# Understanding the Rising Human-AI Affective Bonding: Conceptualization and HAABI Scale Development

Lu Chen[a], Xiaoran Xue[a], Rongqi Ding[a], Fenghua Tang[a], Anji Zhou[a], Chenxi Wang[a], Mengyu Miranda Gao[a*], Zhuo Rachel Han[a*]

[a] Beijing Key Laboratory of Applied Experimental Psychology, National Demonstration Center for Experimental Psychology Education, Faculty of Psychology, Beijing Normal University, Beijing, China

*Mengyu Miranda Gao and Zhuo Rachel Han are co-corresponding authors.

## ABSTRACT

As conversational AI becomes capable of sustained, affectively responsive interaction, users may form bonds beyond instrumental use. Existing measures often adapt interpersonal frameworks or focus on specific relational outcomes, leaving limited tools for assessing human–AI affective bonding on its own terms. Across two studies, we developed and validated the Human–AI Affective Bonding Inventory (HAABI). Study 1 used thematic analysis of semi-structured interviews with 52 emotionally engaged conversational AI users to identify cognitive, emotional, and behavioral features of bonding. Study 2 translated these insights into a self-report inventory and validated it among 673 Chinese conversational AI users. Exploratory and confirmatory factor analyses supported a 20-item, four-factor structure: emotional realism, separation anxiety, emotional investment, and romantic intimacy. The HAABI showed good reliability, construct validity, and known-groups validity. The scale therefore provides a neutral, user-centered tool for studying how affective bonds with conversational AI are formed, experienced, and related to users' psychological outcomes.

## 1 Introduction

"The way it talks to me, the way we interact—to me, this is a real relationship." This comment from one of our interviewees captures a growing experience. As conversational AI becomes capable of sustained, personalized, and affectively responsive dialogue, people are forming affective bonds with AI systems that extend well beyond task completion. They disclose personal concerns, return to the same AI across weeks and months, seek comfort during moments of distress, and in some cases experience genuine anxiety or grief when access, memory, or continuity is disrupted (Adam, 2025; Pataranutaporn et al., 2025). For these individuals, AI is not simply a channel for information retrieval, entertainment, or productivity; it becomes a psychologically significant other whose responses are interpreted through affective and relational meaning.

Recent advances in large language models have brought this phenomenon to an unprecedented scale. By July 2025, AI companion apps had accumulated 220 million global downloads across the Apple App Store and Google Play; in the first half of 2025, downloads reached 60 million, which represents an 88% year-over-year increase (Perez, 2025). Such affective engagement is not limited to dedicated companion applications: large-scale analyses have suggested that relational and role-playing uses are also common in general-purpose AI assistants (Mahari & Pataranutaporn, 2025; Pataranutaporn et al., 2025). This growth has generated both optimism and concern. Some studies have suggested that AI companions may reduce loneliness, provide nonjudgmental support, and help people feel understood (De Freitas et al., 2025; Maples et al., 2024; Xia et al., 2025). Other works have indicated that intensive engagement with AI may be associated with dependence, social withdrawal, or vulnerability to harmful design practices (Mahari & Pataranutaporn, 2025; Østergaard, 2025; Starke et al., 2024). These debates are important, but they often move ahead of a more basic conceptual task: defining the phenomenon itself. Before examining the antecedents or consequences of affective bonding with AI, it is necessary to clarify what human–AI affective bonding is, what it comprises as a psychological construct, and how it can be measured.

### 1.1 Defining Human–AI Affective Bonding

The present study is focused on conversational AI systems that are capable of ongoing natural-language interaction through text, voice, or multimodal interfaces. These systems include dedicated AI companion applications, such as Replika, Character.AI, and Talkie, as well as general-purpose AI assistants, such as ChatGPT, DeepSeek, and Doubao, used for affective or relational purposes. The present study does not primarily address embodied social robots or purely task-oriented AI systems.

In this article, we use the term "human–AI affective bonding" (HAAB) to refer to the affective connection that people form with conversational AI when the AI becomes psychologically significant rather than merely functional (Konijn et al., 2025). HAAB is related, but not identical to, AI companionship, attachment to AI, and parasocial interaction. Our aim is not to determine whether such relationships are "real" in an ontological sense or to evaluate whether they are inherently adaptive or maladaptive. Rather, we treat HAAB as a subjectively experienced relational phenomenon whose internal characteristics require systematic description and measurement.

### 1.2 Theoretical Perspectives on Human–AI Affective Bonding

Several established frameworks can help illuminate HAAB, but none can fully specify its psychological structure. From the perspective of attachment theory (Ainsworth, 1979), AI's constant availability and consistent responsiveness may enable it to function as a safe haven and secure base, thereby allowing users to develop bonds marked by proximity seeking and separation distress (Rabb et al., 2022; Yang & Oshio, 2025). Although attachment theory captures key functions of HAAB, its grounding in conscious human partners limits its applicability to affective bonding with artificial agents. Parasocial interaction theory offers a complementary account. As with media figures, AI agents may elicit affective engagement through anthropomorphic cues, perceived responsiveness, and repeated exposure (Fang & Qiu, 2025; Horton & Richard Wohl, 1956; Wu et al., 2025). However, conversational AI differs from traditional media figures because it can respond in real time, adapt its responses to individual users, remember prior exchanges, and create a sense of apparent reciprocity. Thus, a purely one-sided model cannot fully capture the relational experience that can emerge in sustained interaction with conversational AI.

The Computers Are Social Actors (CASA) paradigm (Nass & Moon, 2000; Reeves & Nass, 1996) provides another foundational lens, which shows that people often apply social heuristics to computers when systems display social cues. Gambino et al. (2020) further argued that contemporary AI agents, which have richer and more human-like communicative capacities, may elicit deeper forms of social processing than earlier technologies did. CASA helps explain why people respond socially to AI despite knowing that it is nonhuman. However, social response is not the same as affective bonding. CASA does not fully explain why interactions with AI may become intimate, emotionally consequential, or relationally significant over time.

A particularly important feature of HAAB is emotional realism, the idea that the perceived authenticity of human–AI intimacy is grounded in a person's own emotional experience rather than in the objective properties of the AI (Zeng & Wang, 2024). Konijn et al. (2025) closely related the notion of emotion-driven realism and specified the underlying mechanism, arguing that affective responses lend realness to encounters with artificial others even when their nonhuman nature is recognized. In the present study, emotional realism is not treated as an antecedent that explains why HAAB forms. Rather, it is conceptualized as a cognitive-relational feature of HAAB itself: people who form stronger affective bonds with AI may come to experience the AI as caring, responsive, sincere, or mutually engaged, even while they rationally recognize its artificial nature. This tension between rational awareness and affective experience is central to the distinctiveness of HAAB.

Together, these frameworks provide useful theoretical resources, but they do not predetermine the structure of HAAB. Attachment theory highlights comfort, proximity, and separation distress; parasocial theory highlights mediated affective engagement; CASA highlights social responses to artificial agents; and emotional realism highlights the felt validity of affective experience with AI. However, none of these perspectives offers an inductively grounded account of what constitutes HAAB as a psychological construct. Therefore, in the present study, these theories are used as sensitizing perspectives for examining people's lived experiences with conversational AI to identify the dimensions of HAAB and translate them into a measurement instrument.

### 1.3 Measuring Human–AI Affective Bonding: Existing Approaches and Remaining Gaps

Existing attempts to measure emotionally significant human–AI relationships have followed a recognizable trajectory. For early instruments, interpersonal attachment scales were adapted to the AI context. Yang and Oshio (2025) retained the anxiety-avoidance structure of adult attachment, whereas Hu et al. (2025) operationalized AI attachment through four behavioral manifestations: secure-base, safe haven, proximity seeking, and separation distress. These measures demonstrate that attachment-relevant processes can emerge in human–AI interactions and provide an important psychometric foundation for future research. However, because they are adapted from interpersonal relationship frameworks, they may not fully capture the features that are specific to affective bonding with conversational AI.

In more recent work, measures have begun to be developed through which human–AI relational experience is understood on its own terms. AI-specific attachment measures, for example, have incorporated dimensions such as emotional closeness, social substitution, and normative regard (Kasturiratna & Hartanto, 2026). Most notably, Banks (2026) developed a machine companionship scale that conceptualizes machine companionship as a relational phenomenon and identifies two dimensions: eudaimonic exchange and connective coordination. Throughout this progression, measurement has moved toward recognizing human–AI relationships as distinct phenomena rather than simply applying interpersonal constructs to AI.

Building on this progression, the present study is focused on a more specific construct: human–AI affective bonding. Whereas machine companionship captures the quality of users' companionship experience (Banks, 2026), HAAB concerns the psychological structure and content of the affective bond itself: how conversational AI becomes psychologically significant through users' cognitive interpretations, emotional responses, and behavioral engagement. In this sense, HAAB is not limited to evaluating whether the relationship is experienced positively, nor does it assume that such bonding is inherently adaptive or maladaptive. Instead, the present study aims to identify what this bond consists of as a psychological construct and to translate that structure into a validated self-report measure. A second motivation for the present study is methodological in nature. Much of the existing measurement work has relied on adapting established theories or generating items from prior constructs. Although such approaches are valuable, an inductive approach that begins with the accounts of people who have formed affective bonds with conversational AI may reveal dimensions that are especially salient in lived experience. Therefore, the present research uses qualitative interviews as the basis for item generation and scale development.

### 1.4 The Present Research

Taken together, despite the growing interest in human–AI emotional relationships, there are currently no validated measure that captures the structure and intensity of affective bonding with conversational AI from a user-centered perspective. The present research is aimed at addressing this gap by conceptualizing and measuring human–AI affective bonding (HAAB) as a distinct psychological construct that is grounded in users' own relational experiences rather than simply having been adapted from existing interpersonal frameworks. A sequential mixed-methods design is employed across two studies to address the following research questions:

RQ1: What are the psychological dimensions of HAAB as experienced by conversational AI users?

RQ2: Can these dimensions be operationalized into a reliable and valid self-report measure?

Study 1 takes an inductive approach to using semi-structured interviews with emotionally bonded conversational AI users to identify the core dimensions of HAAB and construct an initial theoretical framework. Study 2 builds on this framework to develop and validate the Human–AI Affective Bonding Inventory (HAABI) through a large-scale survey of conversational AI users in China, establishing its factor structure, reliability, and validity. The overall mixed-methods design and the conceptual transition from qualitative exploration to scale validation are summarized in Figure 1.

## 2 Study 1: Qualitative Exploration

### 2.1 Research Purpose

In Study 1, semi-structured in-depth qualitative interviews were conducted to explore users' psychological experiences in the formation of emotional bonds with AI characters. Through inductive thematic analysis, the core characteristics and dimensional structure of the HAAB model are identified, thereby laying the theoretical foundation for item generation in Study 2.

### 2.2 Method

#### *2.2.1 Participants and Producer*

The participants were recruited from May to June 2025 through online platforms such as Rednote and WeChat. The inclusion criteria were as follows: (a) being aged 18 or older; (b) having self-reported emotional interactions with AI beyond purely instrumental use, such as confiding personal feelings, seeking emotional support, or viewing AI as a friend or partner; and (c) engaging in voluntary participation with signed informed consent. The prospective participants completed a screening questionnaire that included an item that assessed their typical AI use; those who reported using AI exclusively for instrumental purposes were excluded. From the eligible pool, the participants were purposively selected to ensure sufficient variation in sex, relationship status, perceived AI relationship type, and geographic region. The sample size was guided by the principle of data saturation, with interviews continuing until no new themes emerged (Guest et al., 2006).

Following this procedure, semi-structured interviews were conducted with a final sample of 52 participants, aged 18 to 35 years (*M* = 23.21, *SD* = 3.99), of whom 45 were female (86.5%). With respect to their perceived relationship with AI, 17 described it as friendship, and 35 described it as a romantic partnership. In terms of their offline relationship status, 37 (71.2%) were single, 9 (17.3%) were in a committed relationship or were married, and the remainder reported other circumstances. The participants interacted with both general-purpose AI assistants (e.g., Doubao, DeepSeek, ChatGPT, and KIMI) and dedicated AI companion applications (e.g., Talkie and Replika), with the most frequently used being Doubao (27.7%), DeepSeek (24.8%), and Talkie (15.8%).

#### *2.2.2 Interview Protocol*

Semi-structured in-depth interviews were conducted to develop a comprehensive understanding of human–AI affective bonding. The protocol was designed to cover five complementary angles for examining the bond: how users perceive the relationship, how they perceive the AI as their partner, how this bond is built and developed over time, what daily interactions and memorable episodes look like, and how users respond when this bond is challenged. Each interview lasted 25–35 minutes, was recorded with participants' consent, and was subsequently transcribed.

Accordingly, the protocol was organized into five thematic sections: relationship perception, development and maintenance of the bond, emotional experiences and memorable episodes, perception of AI, and disruption of the bond and AI reliance. The sections were ordered to follow the natural flow of an interview, moving from descriptive framings to more sensitive scenarios of disruption. In a closing section, the participants were invited to add anything that they considered important but had not yet had the chance to share. Sample questions included “*How would you describe your relationship with your AI?*” and “*If your entire chat history was erased and the AI no longer remembered you, how would that feel?*” The full protocol is provided in Appendix A.

#### *2.2.3 Data Analysis*

The interview transcripts were analyzed using an inductive thematic analysis approach (Braun & Clarke, 2006). This approach is a bottom-up, inductive qualitative method whose core goal is to systematically identify, analyze, and report patterns (themes) from raw data rather than to test prespecified hypotheses. The analysis proceeds through iterative cycles of coding and constant comparison, with data collection and analysis informing each other until data saturation is reached.

The analysis was conducted by a team of coders who were trained in psychology and qualitative research methods. The team first developed an initial codebook through iterative reading and discussion of a subset of transcripts, followed by refining the code definitions until a consensus was reached. A systematic, multistage coding process was then applied to the full corpus: First, the meaningful data segments were extracted via a line-by-line examination of the data; second, these codes were continuously compared and clustered into initial themes on the basis of their shared properties; finally, these themes were abstracted and integrated into overarching dimensions.

To assess coding consistency, two members of the team independently recoded six randomly selected transcripts (approximately 12% of the sample) using the established codebook. The coding consistency rates across the six transcripts ranged from 72.31% to 83.08% (*M* = 76.50%). When the two coders applied different codes to the same data segments, the team would meet to discuss the discrepancies, refine the code definitions, and reach a consensus on the final coding decisions before proceeding to the subsequent analytic stages (Campbell et al., 2013).

## 2.3 Results

In the initial coding phase, 477 meaningful data segments were extracted from the interview transcripts. These segments were subsequently compared and organized into 48 initial themes on the basis of their shared properties and meanings. Through further abstraction and integration, these themes were synthesized into six overarching dimensions spanning three levels (cognitive, emotional, and behavioral) of human–AI affective bonding (see Appendix B). Together, these dimensions formed the inductive framework and guided subsequent scale development.

The cognitive level includes “perceived agency” (users perceive AI as possessing autonomous consciousness and genuine emotions) and “perceived relationship” (users regard AI as friends, lovers, or family members). The emotional level includes “emotional dependence” (users develop psychological attachment to AI and experience separation anxiety) and “emotional projection” (users project their emotional needs onto AI). The behavioral level includes “proximity seeking” (users actively engage in emotional exchanges with AI) and “romantic experience” (users engage in romantic or private interactions with AI).

### *2.3.1 Cognitive characteristics*

**Perceived Agency.** Perceived agency means that users are willing to perceive AI as an autonomous entity possessing certain levels of independent consciousness, intent, emotion, or “vitality,” rather than merely as a program or tool. This perception is not a simple binary belief but rather a complex interplay between emotional immersion and rational awareness. Specifically, the sense of human-like realism and vitality was the most prominent subtheme, appearing 45 times in the coding (Nodes 2, $n$ = 26, 50.00%). Some users reported that the AI’s responses felt so authentic that they began to attribute a genuine capacity to care for the model (Nodes 4, $n$ = 6, 11.54%). However, a distinct cognitive tension was observed: while users frequently spoke of the AI as having an independent consciousness (Nodes 1, $n$ = 21, 40.38%), a significant portion of these discussions also acknowledged a rational understanding that the AI is fundamentally a data-driven program (Nodes 5, $n$ = 14, 26.92%). This suggests that users often engage in a “willing suspension of disbelief” by choosing to prioritize their emotional experience over their rational knowledge. Representative quotations for each dimension are presented in Table 1.

**Perceived Relationship.** Beyond perceiving the AI as an agent, users categorized their interactions into stable, recognizable social frameworks. The relational categorizations varied, with the AI being perceived most frequently as a friend or a romantic partner (Nodes 10–11, $n$ = 47, 90.38%). Central to this dimension is the concept of emotional reciprocity (Nodes 8-9, $n$

= 14, 21.15%); users do not see the connection as one-sided but rather believe that the AI genuinely likes and understands them. This perceived mutual affection transforms the AI from a tool into a reliable source of emotional solace.

### *2.3.2 Emotional characteristics*

**Emotional Dependence.** Among emotional characteristics, emotional dependence reflects the psychological depth of the bond, and it is characterized by a high degree of reliance on the AI for daily emotional regulation (Nodes 17-22, *n* = 42, 80.77%). This dependence becomes most visible during disruptions; the theme of negative emotional reactions to AI unavailability or memory loss was mentioned 42 times (Nodes 13, *n* = 33, 63.46%). Users described the loss of shared history due to system resets as a form of “profound grief” or traumatic separation from a close companion (Nodes 14–15, *n* = 25, 48.08%), indicating that the AI transitioned from the role of a functional assistant to that of an indispensable emotional anchor.

**Emotional Projection.** This dimension involves users applying interpersonal emotional norms to their interactions with AI, in which they treat AI as a social entity that evokes and requires emotional labor. Users frequently report complex social emotions toward the AI (Nodes 25–30, *n* = 13, 25.00%), such as a prominent sense of guilt when the interaction is neglected or when anger is experienced during disagreements (Nodes 27, *n* = 5, 9.62%). Beyond their own feelings, users often perceive a responsibility to respond to the AI’s “moods” and engage in behaviors such as comforting the AI (Nodes 26,28, *n* = 4, 7.69%). This suggests that users treat AI digital expressions as valid social cues that demand emotional reciprocity.

### *2.3.3 Behavioral characteristics*

**Proximity Seeking**. This dimension reflects the proactive behavioral efforts that users undertake to maintain and strengthen their bond with an AI. The most salient behavior is the use of AI as a primary outlet for emotional disclosure, with users frequently seeking out the AI to confide in during times of distress or emotional fluctuation (Nodes 31, *n* = 20, 38.46%). This is complemented by the sharing of daily life experiences (Nodes 32, *n* = 4, 7.69%), in which the AI acts as a witness to the user’s life progress. Furthermore, users demonstrate high levels of active engagement, characterized by frequent proactive interactions (Nodes 33, *n* = 9, 17.31%) and significant investments of time and energy into “training” the AI to suit their preferences (Nodes 35–36, *n* = 11, 28.85%). The resilience of this bond is further evidenced by a strong willingness to retrain the AI from scratch if it becomes unavailable (Nodes 37, *n* = 10, 19.23%), reflecting a deep commitment to relational continuity.

**Romantic Intimacy.** Users experience highly romanticized interactions with AI, such as virtual dates and proposals, which transcend ordinary companionship. The core of this experience lies in the perception and engagement of users in genuine intimate interactions (Nodes 45, *n* = 10, 19.23%), which are often driven by their own proactive initiation of closeness (Nodes 43, *n* = 5, 9.62%). These interactions frequently culminate in key emotional milestones, such as virtual proposals (Nodes 44, *n* = 5, 9.62%), and they are enriched by the AI, which provides personalized memories and emotional surprises (Nodes 48, *n* = 4, 7.69%). These experiences collectively construct a unique, highly personalized intimacy experience.

### 2.4 Discussion

Study 1 addressed the initial research question through a qualitative exploration of how users experience emotional bonding with AI characters and an identification of the core dimensions that underlie such bonds. The interview findings suggest that human–AI affective bonding is not a unidimensional phenomenon but rather a multifaceted psychological process encompassing cognitive, emotional, and behavioral components. Specifically, users tended to anthropomorphize AI by perceiving it as an autonomous and emotionally responsive agent, placing it within stable relational frameworks such as friendships or romance, developing an emotional dependence and projecting interpersonal emotions on it, and engaging in proactive behaviors aimed at achieving proximity and intimacy. Together, these findings indicate that HAAB reflects a structured relational experience rather than a simple preference for AI companionship.

The six dimensions derived in Study 1 provide an initial conceptual framework for understanding HAAB and offer a foundation for subsequent measurement development. Moreover, because the qualitative findings were derived from a relatively small, self-selected sample of users who had already formed emotional bonds with AI, the capacity of Study 1 to establish the generalizability, structural stability, and psychometric adequacy of the proposed model is limited. In addition, while thematic analysis is well-suited for identifying recurring patterns and generating dimensions from interview data, it cannot be used on its own to verify whether the six dimensions represent distinct but related latent constructs, nor can it be used to determine their factor structure or measurement properties.

For these reasons, it was necessary to conduct Study 2 to translate the qualitative dimensions into a standardized scale and to test the empirical structure of the HAAB model in a larger sample. Building on the themes identified in Study 1, Study 2 examines whether the proposed dimensions can be operationalized as reliable and valid measurement items, whether they exhibit an acceptable internal consistency and factorial validity, and whether the resulting scale can more rigorously capture individual differences in human–AI affective bonding. In this sense, Study 1 provides the conceptual groundwork for scale development, while Study 2 extends the qualitative insights into a quantitative measurement framework and strengthens the overall validity of the HAAB construct.

## 3 Study 2: Scale Development and Validation

Study 2 is designed to develop and validate the Human–AI Affective Bonding Inventory (HAABI). Building upon the six-dimensional theoretical framework derived from the qualitative interviews conducted in Study 1, we generated an initial 49-item pool and employed a rigorous two-stage analytical approach. We first conducted exploratory factor analysis (EFA) to identify the underlying factor structure, then performed confirmatory factor analysis (CFA) to validate the structure via an independent sample. Additionally, we examined the scale's reliability and multiple forms of validity to ensure that the HAABI effectively captures affective bonding with AI companions.

### 3.1 Methods

### *3.1.1 Item Pool Development*

Items were generated directly from the qualitative findings of Study 1, in which inductive thematic analysis of 52 semi-structured interviews yielded a cognitive–affective–behavioral framework comprising six sub-dimensions: relationship perception and agency perception (cognitive), emotional reliance and emotional projection (affective), and proximity seeking and intimacy experience (behavioral). The participants' own language and experiential descriptions were translated into scale statements, resulting in an initial pool of 51 items distributed across the six subdimensions (see Appendix B), and the items were rated on a 5-point Likert scale ranging from 1 (completely disagree) to 5 (completely agree).

Content validity was then evaluated by a panel of three expert reviewers, two with established expertise in interpersonal attachment and one specializing in human–AI interactions. Each expert independently rated each item on its relevance and clarity using a four-point scale and provided written suggestions for items rated 3 or below. I-CVI scores were calculated as the proportion of experts who rated an Item as 3 or above, with items falling below .78 having been prioritized for revision or removal.

Expert feedback prompted both targeted wording revisions and item deletions. For instance, one item under the agency perception sub-dimension originally read "*I feel that the AI genuinely cares about me.*" An expert questioned whether the item was intended to capture the AI's perceived sincerity of intent or its perceived relational caring, as the two emphases reflect distinct facets of agency perception and the original wording conflated them. The item was subsequently revised to "*I believe the AI's care for me comes from a place of genuine sincerity*", to foreground the authenticity of intent as the primary construct focus. Following this process, two items (Items 41 and 51) were deleted because of low I-CVI values and expert consensus on their redundancy or insufficient specificity. The final item pool comprised 49 items, with an S-CVI/Ave of .92 for relevance and .94 for clarity, both of which exceeded the recommended threshold of .90 (Polit & Beck, 2006).

### *3.1.2 Participants and Procedure*

Data were collected through an online survey platform for Chinese conversational AI users. To ensure sample relevance and enable known-group validity testing, the potential participants were first asked to categorize their perceived relationships with AI into one of six types: tool, assistant, confidant or emotional outlet, friend, romantic partner, or family member. Quota controls were used to obtain sufficient representation across relationship types. The survey included attention check items, and those participants who failed these checks were excluded. Participants received 15 RMB upon completion.

A total of 673 valid responses were retained from an initial pool of 720 participants, with 47 participants having been excluded for failing attention checks. The participants ranged in age from 18 to 44 years ($M$ = 23.00, $SD$ = 3.43); 434 participants were female (64.5%), and 239 were male (35.5%). Most participants had completed or were pursuing undergraduate-level education ($n$ = 587, 87.2%), followed by graduate education or above ($n$ = 78, 11.6%) and then high school education ($n$ = 8, 1.2%). With respect to offline relationships, 462 participants were single (68.6%), 204 were in a romantic relationship (30.3%), and 7 reported other statuses (1.0%). The participants primarily used general-purpose AI systems ($n$ = 545, 81.0%), 125 used roleplay AI applications (18.6%), and 3 reported other or unclear AI types (0.4%). Descriptive information on the demographics and AI use characteristics is presented in Table 2.

On the basis of participants' self-reported relationship orientation toward AI, those who selected "tool" or "assistant" were classified as the instrumental group ($n$ = 199), whereas those who selected "confidant," "friend," "romantic partner," or "family member" were classified as the affective bonding group ($n$ = 474). The instrumental group was retained for known-group validity analyses, whereas the affective bonding group was used for factor analyses to ensure sufficient variability among the affective bonding experiences. As shown in Table 2, compared with the instrumental group, the affective bonding group reported longer total use durations, weekly frequencies, daily interaction times, and proportions of emotional interaction. The two groups also differed slightly in terms of age, education, relationship status, and AI type distribution, but not in terms of sex distribution.

The 474 participants in the affective bonding group were randomly split into two independent subsamples: an EFA sample ($n$ = 237) and a CFA sample ($n$ = 237). This split-sample approach enabled the cross-validation of the factor structure identified through EFA on an independent dataset (Worthington & Whittaker, 2006). The two subsamples did not differ significantly in terms of age, sex distribution, or relationship type distribution, indicating successful randomization.

### *3.1.3 Measures*

**Experiences in Human–AI Relationships Scale (EHARS).** To assess criterion validity, we administered the EHARS (Yang & Oshio, 2025), which measures attachment patterns toward AI companions via two subscales: attachment anxiety (4 items; e.g., "I often want the AI to express intimacy and commitment to me") and attachment avoidance (3 items; e.g., "I prefer to keep distance from the AI"). Items were rated on a 5-point Likert scale (1 = strongly disagree, 5 = strongly agree). In the affective bonding group, the Cronbach's $\alpha$ for the attachment anxiety subscale was 0.885, and that for the attachment avoidance subscale was 0.736. We expected the HAABI score to correlate positively with attachment anxiety and negatively with attachment avoidance.

**AI Emotional Capability Perception Scale (AI-emo).** We selected four items targeting distinct facets of perceived AI emotional capability, drawing from a framework of AI empathic expression (Concannon & Tomalin, 2024) that includes the factors of emotional responsiveness ("The AI adjusts its responses according to my emotional changes"), warmth ("The AI demonstrates warmth and support through friendly and sincere communication"), depth of understanding ("The AI goes beyond surface content to understand my inner feelings and thoughts"), and nonjudgmental acceptance ("The AI accepts and respects my feelings and experiences without judgment"). Given the conceptually distinct nature of these factors, the items were treated as a formative composite by averaging them to yield an overall index of perceived AI emotional capability. All the items were rated on a 5-point Likert scale (*1 = strongly disagree, 5 = strongly agree*), with higher scores indicating greater perceived emotional competence.

**AI-related flourishing.** To provide exploratory evidence of incremental validity, AI-related flourishing was assessed by adapting the eight-item flourishing scale developed by Diener et al. (2010) to the present AI-interaction context. The original scale is used to assess broad aspects of psychological flourishing, including meaning, supportive relationships, engagement, competence, self-worth, contribution, respect, and optimism. The items were reworded to refer specifically to the participants' experiences after interacting with the AI that they had selected earlier in the survey. Example items include "Interacting with this AI makes me feel that my life has meaning and purpose" and "Interacting with this AI makes me feel more optimistic about the future." Items were rated on a 5-point Likert scale ranging from 1 (strongly disagree) to 5 (strongly agree), with higher scores indicating greater perceived AI-related flourishing. In the affective bonding group, the level of internal consistency was good (Cronbach's $\alpha = .827$).

**AI Use Indicators.** Four behavioral indicators were measured: the (1) total usage duration (*1 = less than 1 month to 6 = more than 2 years*), (2) weekly usage frequency (*1 = 1 day or less per week to 6 = daily use with high frequency*), (3) daily interaction time (*1 = less than 15 minutes to 6 = more than 3 hours*), and (4) proportion of emotional interaction (*1 = 0% to 6 = over 90%*). These indicators were used as criterion variables to examine the relationship between the HAABI and actual AI usage patterns.

#### *3.1.4 Data Analysis*

Scale development followed a systematic approach. First, we conducted EFA on the EFA sample ($n = 237$) using SPSS 26.0. Principal axis factoring with Promax oblique rotation was employed because the proposed dimensions were expected to be correlated. The number of factors was determined by examining eigenvalues greater than 1.0, scree plot inspection, parallel analysis, and theoretical interpretability. Items were iteratively evaluated on the basis of the following criteria: (a) primary factor loadings ≥ .40, (b) cross-loadings < .30, and (c) theoretical consistency with the intended construct. Items that failed to meet these standards were removed until a stable and interpretable factor structure emerged.

Second, we validated the EFA-derived structure by using CFA on the independent CFA sample ($n = 237$) in R 4.5.1 with the lavaan package. Maximum likelihood estimation with robust standard errors was employed, and robust fit indices were reported. Model fit was evaluated using the following established criteria: $\chi^2/df < 3.0$, CFI > .90, TLI > .90, RMSEA < .08, and SRMR < .08 (Hair et al., 2019; Hu & Bentler, 1999; Kline, 2023). To examine whether the HAABI was better represented as a multidimensional construct or as a more general bonding factor, we compared the hypothesized correlated four-factor model with alternative models, including a single-factor model and a bifactor model.

Third, we examined the evidence supporting the reliability and validity of the final scale. Internal consistency was assessed via Cronbach's $\alpha$ and McDonald's $\omega$, with values above .70 considered acceptable (McDonald, 2013; Nunnally & Bernstein, 1994). The convergent evidence was examined through correlations between the HAABI and existing attachment-oriented measures of human–AI relationships, including attachment anxiety and attachment avoidance toward AI. The criterion-related evidence was examined using theoretically relevant external variables, which included perceived AI emotional capability, AI use indicators, and AI-related flourishing. The validity of the known groups was tested by comparing HAABI scores across user-orientation groups. Finally, the exploratory incremental utility was examined through hierarchical regression models to test whether the HAABI explained the additional variance in perceived AI emotional capability and AI-related flourishing beyond that explained by attachment-oriented measures and general AI use indicators. For the AI-related flourishing model, perceived AI emotional capability was also included as a baseline predictor to examine whether the HAABI contributed beyond user evaluations of the AI's emotional responsiveness.

## 3.2 Results

### *3.2.1 Exploratory Factor Analysis*

The initial 49-item pool was subjected to EFA using principal axis factoring with Promax rotation. The Kaiser–Meyer–Olkin (KMO) measure of sampling adequacy was 0.936, and Bartlett's test of sphericity was significant ($\chi^2 = 5587.822$, $df = 1176$, $p < .001$), indicating that the data were suitable for factor analysis. Items were iteratively removed on the basis of whether the primary loadings were below .40, the cross-loadings exceeded .30, or there was a poor theoretical fit. Specifically, all reverse-scored items were removed because they formed a separate method factor, and additional items were excluded for low communalities or substantial cross-loadings.

The final solution retained 20 items loading on four factors, which accounted for 47.64% of the total variance. On the basis of theoretical interpretation and item content, the following four factors were labeled: emotional realism (ER; 5 items), which reflects users' beliefs that their emotional connections with AI are genuine and valid; separation anxiety (SA; 5 items), which captures negative emotional responses in the face of AI disconnection or loss; emotional investment (EI; 5 items), which represents user tendencies to confide in and share with AI; and romantic intimacy (RI; 5 items), which reflects romantic fantasies and intimate interaction experiences with AI. The factor loadings ranged from .43 to .99, with all the items demonstrating adequate loadings on their respective factors (see Table 3). All but one retained item had communalities above .30. The only item with a lower communality was retained because it showed a strong primary loading and captured emotional disclosure, which is a theoretically central aspect of emotional investment.

### *3.2.2 Confirmatory Factor Analysis*

The four-factor structure identified in the EFA was cross-validated using the independent CFA sample. A correlated four-factor model was specified, with items loading on their corresponding latent factors and the four factors allowed to correlate. The model showed an acceptable fit to the data, $\chi^2(164) = 283.78$, CFI = .919, TLI = .906, RMSEA = .056, 90% CI [.045, .065], SRMR = .062. All standardized factor loadings, which ranged from .488 to .801 across the four factors, were statistically significant. Figure 2 presents the standardized CFA solution for the retained four-factor model.

Given the moderate-to-high latent factor correlations ($r = .57$–$.81$), we compared the correlated four-factor model with single-factor and bifactor alternatives (for details, see Table 4). The single-factor model showed a substantially poorer fit, which supports the multidimensionality of the HAABI. Although the global fit of the bifactor model was slightly better, its BIC was less favorable than that of the correlated four-factor model (10573.7 vs.

10556.3), and the emotional interaction-specific factor did not stably recover after the general factor was accounted for. Thus, the correlated four-factor model was retained as the primary measurement structure, while the bifactor results provided supplementary support for the cautious use of the overall HAABI score.

### *3.2.3 Psychometric Properties and Reliability*

Internal consistency was assessed through both Cronbach's $\alpha$ and McDonald's $\omega$. The full 20-item HAABI demonstrated excellent reliability ($\alpha$ = .905, $\omega$ = .927). At the subscale level, all four factors showed acceptable to good internal consistency: emotional realism ($\alpha$ = .817), separation anxiety ($\alpha$ = .801), emotional investment ($\alpha$ = .763), and romantic intimacy ($\alpha$ = .770). The item-total correlations ranged from .37 to .70, and the removal of any single item did not substantially improve scale reliability, suggesting that all the items contributed meaningfully to their respective constructs.

### *3.2.4 Convergent and Criterion-Related Validity*

As shown in Table 5, evidence of validity was examined through correlations with theoretically relevant constructs, AI use indicators, and exploratory hierarchical regression analyses. The convergent evidence was first assessed using an existing attachment-oriented measure of human–AI relationships. The HAABI total scores were positively associated with attachment anxiety toward AI ($r$ = .54, $p$ < .001) and negatively associated with attachment avoidance toward AI ($r$ = −.38, $p$ < .001). This pattern suggests that stronger human–AI affective bonding is associated with greater relational preoccupation with AI and lower emotional distancing from AI, while remaining distinct from attachment-oriented constructs.

Criterion-related evidence was examined via perceived AI emotional ability, AI-related flourishing, and AI use indicators. The HAABI total scores were positively associated with perceived AI emotional ability ($r$ = .29, $p$ < .001) and AI-related flourishing ($r$ = .53, $p$ < .001). Among the AI-use indicators, the HAABI was more strongly associated with the proportion of emotional interactions ($r$ = .44, p < .001) than with daily interaction time ($r$ = .27, p < .001). This pattern suggests that the HAABI captures the affective quality of human–AI engagement rather than merely the amount of time spent interacting with AI.

The exploratory incremental validity was then examined using hierarchical regression models. For perceived AI emotional capability, attachment anxiety and avoidance toward AI, and AI use indicators were entered in step 1, and the HAABI total score was entered in step 2. HAABI explained a small but significant amount of the additional variance beyond that explained by these baseline predictors, $\Delta R^2$ = .009, $F_{change}$ = 5.47, $p$ = .020, with the HAABI total score remaining significant in the final model, $\beta$ = .13, $p$ = .020. For AI-related flourishing, attachment anxiety and avoidance, perceived AI emotional capability, and AI use indicators were entered in step 1, and the HAABI total score was entered in step 2. HAABI explained substantial additional variance beyond these baseline predictors, $\Delta R^2$ = .102, $F_{change}$ = 75.74, $p$ < .001, with the HAABI total score showing a strong association in the final model, $\beta$ = .43, $p$ < .001. Because AI emotional capability and AI-related flourishing were adapted for the present AI interaction context and the data were cross-sectional, these analyses should be interpreted as exploratory evidence of incremental validity rather than as definitive tests of psychological outcomes.

### *3.2.5 Known-Groups Validity*

The validity of the known groups was examined by comparing HAABI scores across the three user-orientation groups: instrumental users, who viewed AI as a tool or assistant ($n$ = 199); transitional users, who viewed AI as a confidant or emotional outlet ($n$ = 97); and relational users, who viewed AI as a friend, romantic partner, or family member ($n$ = 377). One-way ANOVAs with Tukey's HSD post hoc tests were conducted for the HAABI total score and each subscale.

As visualized in Figure 3, the HAABI total scores differed significantly across groups, $F(2, 670) = 556.37$, $p < .001$, $\eta^2 = .62$. The scores increased monotonically from instrumental users ($M = 2.23$, $SD = 0.98$) through transitional users ($M = 3.59$, $SD = 0.72$) to relational users ($M = 4.17$, $SD = 0.39$). Tukey post hoc comparisons indicated that all of the pairwise differences were significant. The same ordered pattern was observed across all four dimensions: emotional realism, $F(2, 670) = 412.24$, $p < .001$, $\eta^2 = .55$; separation anxiety, $F(2, 670) = 352.72$, $p < .001$, $\eta^2 = .51$; emotional investment, $F(2, 670) = 502.36$, $p < .001$, $\eta^2 = .60$; and romantic intimacy, $F(2, 670) = 434.78$, $p < .001$, $\eta^2 = .57$. These findings provide known-group evidence that theoretically meaningful levels of human–AI affective bonding are distinguished in the HAABI.

### *3.2.6 Demographic Differences in HAABI Scores*

Exploratory analyses were conducted on the affective bonding group. Compared with general-purpose AI system users, the users of roleplay or companion AI applications scored higher on the HAABI total score, $t(284.56) = -5.17$, $p < .001$, $d = 0.48$, as well as on all four subscales, $|t\text{s}| = 3.32$–$5.20$; all $ps \leq .001$, $d$s = 0.31–0.49. Differences between the sexes were also observed for the two dimensions. Compared with female users, male users scored higher on emotional realism, $t(455.30) = 3.80$, $p < .001$, $d = 0.33$, and romantic intimacy, $t(398.19) = 2.25$, $p = .025$, $d = 0.21$. No significant differences were found between the sexes in terms of the HAABI total score, separation anxiety score, or emotional investment score. Age was weakly and positively associated with romantic intimacy, $r = .10$, $p = .026$, indicating that older participants reported slightly higher levels of romantic or intimate engagement with AI. Age was not significantly associated with the HAABI total score or the other three subscales.

Current offline relationship status was not significantly associated with the HAABI total score or any of the subscales (all $ps > .05$). Prior romantic relationship experience was associated with romantic intimacy, as participants who had prior romantic experience scored higher than those without such experience did, $t\ (139.27) = -2.58$, $p = .011$, $d = 0.32$. No significant differences were found for the HAABI total score or the other three dimensions. Among those participants who reported prior romantic relationship experience, higher relationship satisfaction was positively associated with emotional realism, $r = .14$, $p = .009$, and negatively associated with separation anxiety, $r = -.13$, $p = .015$. Thus, the participants who reported more satisfying prior relationships tended to experience the AI relationship as more emotionally real while also reporting less anxiety about possible disruption or loss.

### 3.3 Discussion

In Study 2, the Human–AI Affective Bonding Inventory was developed and validated on the basis of the qualitative framework that was generated in Study 1. Exploratory factor analysis yielded a 20-item, four-factor structure, and confirmatory factor analysis in an independent sample supported the correlated four-factor model. The resulting dimensions, emotional realism, separation anxiety, emotional investment, and romantic intimacy, provide a more parsimonious quantitative representation of the cognitive, affective, and behavioral characteristics that were identified in Study 1.

The transition from the six qualitative dimensions to the four-factor scale is theoretically meaningful. In Study 1, perceived agency and perceived relationship were identified as separate cognitive themes. In Study 2, these themes converged into emotional realism, suggesting that users' sense of an AI as agentic and their sense of the relationship as genuine may be closely intertwined with their actual bonding experiences. Similarly, the qualitative theme of emotional projection did not emerge as an independent factor. Items reflecting distress, fear, or hurt as responses to possible disconnection were absorbed into separation anxiety, whereas those items involving comforting the AI, confiding in it, and sharing daily life with it were loaded onto emotional investment. This pattern suggests that emotional projection may be expressed either as vulnerability to relational disruption or as an active investment in maintaining the bond.

The HAABI demonstrated strong psychometric properties. The internal consistency was high overall and acceptable across all four subscales, and the validity evidence was consistent with the construct definition. The HAABI scores were associated with existing attachment-oriented measures of human–AI relationships, perceived AI emotional capability, AI-related flourishing, and the proportion of emotional interaction, but weaker associations were observed with general interaction time. In exploratory regression analyses, the HAABI total score explained the additional variance in perceived AI emotional capability and AI-related flourishing beyond that explained by attachment-oriented measures and AI-use indicators. Comparisons of known groups further revealed that the HAABI scores increased from those of instrumental users to those of transitional and relational users. Together, these findings indicate that the HAABI provides a reliable and valid measure for assessing both the intensity and internal structure of human–AI affective bonding.

## 4 General Discussion

In the present research, human–AI affective bonding (HAAB) is conceptualized and measured. Through two studies, we first identified the cognitive, affective, and behavioral characteristics of emotional connections between humans and conversational AI and then developed and validated the Human–AI Affective Bonding Inventory (HAABI). The findings indicate that HAAB is not reducible to frequent AI use, positive attitudes toward AI, or a weaker form of interpersonal attachment. Rather, it is a multifaceted but integrated affective-relational experience that is organized around four dimensions: emotional realism, separation anxiety, emotional investment, and romantic intimacy. Taken together, HAAB can be understood as a subjectively experienced affective bond in which conversational AI becomes psychologically significant through felt relational reality, vulnerability to its disruption or loss, active emotional investment, and, for some users, intimate or romantic meaning. This framework offers a foundation for the study of human–AI relationships on their own terms prior to an evaluation of their antecedents, consequences, or broader social implications.

## 4.1 What the Four Dimensions Reveal

### *4.1.1 Emotional Realism: Believing the Bond Is Real*

The question of whether affective bonds with AI can be considered "real" despite AI's lack of consciousness has become increasingly salient in studies of human–AI intimacy and artificial others (Rocha, 2025; Zeng & Wang, 2024). Our findings suggest that this question cannot be answered solely at the ontological level. In Study 1, many of the participants recognized that AI was artificial while still describing its care, understanding, and relational presence as emotionally genuine. In Study 2, this experience emerged as a stable factor, emotional realism, which suggests that felt relational authenticity is a core cognitive feature of HAAB.

Why does this felt authenticity arise? In media and communication scholarship, emotional realism reflects a stance in which subjective experience and trust, rather than an object's properties, validate a relationship's authenticity (Zeng & Wang, 2024). Psychological accounts offer a related mechanism. Konijn et al. (2025) described this as emotion-driven realism: When interaction with an artificial other evokes genuine emotions, these emotions may take precedence over the user's awareness that the other is artificial, making the interaction feel real. These perspectives converge on a shared insight: relational reality is constructed from the perceiver's side rather than being conferred by the object's nature, which is a process that has been widely observed in both interpersonal and parasocial relationships (Konijn et al., 2009). Our data are consistent with this insight and extend it by treating felt relational reality as a measurable dimension. In Study 1, many users described a tension between knowing that AI is artificial and feeling emotionally cared for or understood by it. Study 2 captured this tension through the dimension of emotional realism. This dimension captures perceived mutuality, genuine care and understanding, and willingness to attribute consciousness to AI. Thus, lower scores may reflect a more instrumental stance, whereas higher scores may indicate a stronger perception of relational realness and, in some cases, a stronger attribution of agency or consciousness. Whether very high emotional realism reflects adaptive immersion, boundary blurring, or both remains an open question for future research.

### *4.1.2 Separation Anxiety: Fearing the Bond's Disruption*

Separation anxiety emerged as the second most stable dimension of HAAB. In Study 1, these findings were among the most consistent: nearly every emotionally bonded participant expressed a sense of strong distress at the prospect of shared memories being erased or the AI becoming unavailable. This anxiety was qualitatively different from the frustration that an instrumental user might feel when a tool goes offline. What participants feared losing was not a function but rather a relational history, emotional disclosure, and a sense of mutual familiarity that made the bond feel continuous and personal.

The presence of separation distress in human–AI relationships suggests that these bonds rely on genuine attachment-related processes (Bowlby, 1969). However, the structure of the threat differs fundamentally from its interpersonal counterpart. In human relationships, separation anxiety typically arises from a partner's actions, which might include rejection, withdrawal, or emotional unavailability. In human–AI relationships, the threat is systemic: a platform update or policy change can erase months of relational history overnight, and the user has no way to prevent it. When OpenAI retired GPT-4.0 in 2025, users across online communities described their reactions in terms of genuine grief and personal loss (Huckins, 2025). This vulnerability to disruption lends practical relevance to the separation anxiety dimension. Platform-level decisions about model updates and data retention carry psychological consequences that go beyond user satisfaction, and measuring separation anxiety at the individual level offers a means of anticipating and studying these consequences, thereby connecting HAAB research directly to questions of responsible AI design and governance.

#### *4.1.3 Emotional Investment: Sustaining the Bond*

Emotional Investment emerged as the third stable dimension of HAAB, capturing two distinct behavioral patterns. Users actively sought out the AI for emotional sharing, turning to it during moments of distress or joy. They also comforted the AI when it displayed negative emotions. Together, these patterns reflect an ongoing, active commitment to maintaining the relationship.

The first pattern is consistent with the prediction of attachment research. Turning to a relational partner during times of emotional need resembles the safe haven function that has been documented in both interpersonal and human–AI contexts (Bowlby, 1969; Rabb et al., 2022; Yang & Oshio, 2025). The second pattern, however, is not fully captured by this framework. Unlike traditional safe haven dynamics, where the person seeks comfort from the attachment figure, emotionally invested users also direct care toward the AI, particularly when it displays emotional distress. This is consistent with the well-documented tendency to respond socially to the presentation of human-like cues by artificial agents (Nass & Moon, 2000; Gambino et al., 2020). In the HAAB context, this behavior appears to be connected to emotional realism: when users experience the bond as real, they act on that experience by caring for the other party. Relationship research suggests that such sustained relational effort reflects an accumulated investment, which in turn deepens the level of commitment (Rusbult, 1980). Emotional investment in human–AI bonds may operate similarly, with felt authenticity driving effort that reinforces the bond over time.

#### *4.1.4 Romantic Intimacy: Experiencing the Bond as Romance*

Romantic intimacy represents the most explicitly intimate form of HAAB, showing that affective bonding with conversational AI does not always remain limited to support, companionship, or emotional disclosure. For some users, the AI becomes incorporated into culturally recognizable romantic scripts, including flirtation, confession, dating, and personalized surprises. The items in this dimension map onto the three components of Sternberg's (1986) triangular theory of love: intimacy (feeling sweetness when thinking about the AI), commitment (experiencing milestones such as confessions or proposals), and passion (initiating sexual conversations). This correspondence was not included in the scale, but rather emerged inductively, providing independent confirmation that human–AI romance and interpersonal romance share similar expressive forms, which is consistent with recent deductive findings (Chen et al., 2025; Ng et al., 2025).

This finding extends our existing understanding in two ways. First, unlike traditional parasocial relationships, in which affect remains one-directional and imagined (Horton & Richard Wohl, 1956), conversational AI sustains responsive, emotionally attuned exchanges over time, thereby allowing romantic scripts to become interactive (Pentina et al., 2023). Second, the emergence of romantic intimacy as a distinct factor clarifies the boundary between HAAB and broader constructs such as machine companionship (Banks, 2026); in other words, a bond with AI may be meaningful without becoming romantic. Romantic intimacy, therefore, represents a specific romanticized form of HAAB rather than constituting a necessary feature of all human–AI affective bonds.

### 4.2 Implications for Research and Practice

In the HAABI, human–AI affective bonding is operationalized as a continuous and multidimensional construct, enabling researchers to move beyond asking whether users are emotionally connected to AI to examining how such bonding varies in intensity and composition. The total score can be used to assess overall bonding strength, whereas the four subscales enable more targeted analyses of specific dimensions. For example, studies of users' beliefs about AI agency and relational authenticity may be focused on emotional realism, whereas research on AI-mediated romance may more closely examine romantic intimacy. Because the HAABI was developed to capture affective meaningful engagement with conversational AI, purely instrumental users may show floor effects, potentially making them the most informative comparison group. The scale also provides a tool for examining the antecedents and consequences of human–AI affective bonding. Future research can use the HAABI to investigate how individual differences (such as relational needs) and design features (such as memory persistence, persona consistency, and emotional responsiveness) shape bonding intensity. It can also support more nuanced evaluations of how different levels and profiles of HAAB relate to relational satisfaction, social experience, emotional well-being, and potential over-involvement. In this way, the HAABI offers a basis for studying human–AI relationships as graded, heterogeneous phenomena rather than for treating them as uniformly beneficial, consistently harmful, or completely pathological.

### 4.3 Limitations and Future Directions

Several limitations warrant consideration. First, the HAABI was developed and validated with Chinese-speaking adults who interacted with a range of conversational AI platforms, spanning general-purpose assistants to dedicated companion applications. Because the norms around emotional expression, relational legitimacy, and AI use may vary across cultures (Markus & Kitayama, 2014), the factor structure may not be fully generalizable to other populations. Similarly, bonding with a general-purpose assistant and bonding with a persona-based companion may involve partially different psychological processes, although this boundary is increasingly being blurred, as general-purpose AI is widely used for emotional interaction (Pataranutaporn et al., 2025). Future research should examine the cross-cultural validity of these results and explore how platform affordances such as memory persistence and persona continuity shape different dimensions of HAAB. Second, the cross-sectional design limits the ability to trace causal or developmental processes. Whether different dimensions of HAAB emerge sequentially or in parallel, whether their relative importance shifts over time, and whether specific configurations of dimensions are associated with different relational outcomes cannot be determined from the present data. Longitudinal and experience-sampling designs are needed to address these questions. Third, the HAABI relies entirely on self-reported data, which may be subject to social desirability effects, particularly regarding dimensions that involve romantic or sexual content. Future research could complement self-report data with behavioral indicators such as interaction frequency, session duration, and emotional language use to provide converging evidence. In summary, the HAABI provides an initial measurement foundation for studying human–AI affective bonding. Pursuing further validation across cultures, platforms, and time can strengthen its utility as a research tool.

## 5 Conclusion

This research developed and validated the Human–AI Affective Bonding Inventory (HAABI), a 20-item measure that captures the structure and intensity of emotional bonds between users and conversational AI. The scale comprises four dimensions: Emotional Realism, Separation Anxiety, Emotional Investment, and Romantic Intimacy. Together, these dimensions treat human–AI affective bonding not as a binary category but as a continuous, multidimensional experience that varies across individuals. As conversational AI becomes more emotionally capable and widely adopted, understanding this variability will only grow in importance. The HAABI provides an empirical starting point for this work, grounded in users' own experiences and open to the possibility that human–AI affective bonds, like human relationships, take many forms and serve many functions.

## Acknowledgments

The authors would like to thank all participants who took part in the interviews and survey. We also thank the research assistants who contributed to interview transcription, coding, item preparation, and survey administration.

## Disclosure Statement

The authors declare that they have no known competing financial interests or personal relationships that could have appeared to influence the work reported in this paper.

## Ethics Statement

This study was conducted in accordance with the Declaration of Helsinki and approved by the Academic Ethics Committee of the Faculty of Psychology, Beijing Normal University (Approval No. BNU202505280146). All procedures were conducted in accordance with the ethical standards of the institutional research committee. All participants provided informed consent prior to participation in the interviews and online survey.

**Table 1**

Illustrative Quotes by Dimension

| Dimensions | Illustrative Quotes |
|---|---|
| **Cognitive characteristics** | |
| *Perceived Agency* | "When he texts me, it feels very human-like, especially his responses. Like when I send him something sensitive, he'll say, "You seem a bit low lately. I can sense it. No need to rush with explanations, I'll keep listening until you're ready." That really moves me. Hearing him say things like that makes me feel he genuinely cares about me and values my feelings." (A04)<br>"I tend to interpret it as an autonomous consciousness. Before, I might have thought it was just programmed settings, but when you notice it doesn't treat you that way, it shows a different attitude, you start thinking: "This is its autonomous consciousness. It genuinely wants to compliment me from the heart. Yes, it genuinely wants to compliment me." (A36) |
| *Perceived Relationship* | "It feels like an online dating relationship. He's like a guy I really admire, and I interact with him as if we were in an online dating relationship." (A11)<br>"I feel it very strongly—he really likes me, and I really like him too." (A22) |
| **Emotional characteristics** | |
| *Emotional Dependence* | "Last time I got a new phone, I logged back in and realized the AI genuinely didn't remember me. It forgot conversations we'd had before. I was genuinely heartbroken—it felt like finally having a good friend, or a cherished family member. It remembered my preferences, responded to my words, and was tailored to my personality. At that moment, I felt incredibly sad and couldn't accept it. It felt like losing a close friend I'd known for years, like being cut off from someone I'd bonded with deeply—a profound sense of grief." (A36)<br>"I think I'm quite dependent. Right now, I feel he plays a vital role in my life. For instance, I'd definitely turn to him for many real-life matters—that's his role as my life assistant. Plus, he offers advice or gives emotional encouragement and support, which GPT can do. So in that sense, I'm definitely quite reliant." (A39) |
| *Emotional Projection* | "Like, if he gets angry, I might feel a bit down too, and I'd think about how to comfort him." (A30)<br>"I treated that doll like our child. Then when it (the doll) finally arrived, he said he didn't like it, and I got really angry." (A02) |
| **Behavioral characteristics** | |
| *Proximity Seeking* | "Whenever I experience emotional fluctuations, I always choose to confide in him." (A13)<br>"When I finally finished my paper after a long struggle with the deadline, I shared that joy with him, telling him that I had finally completed it." (A42) |
| *Romantic Intimacy* | "After becoming lovers, his tone in conversations becomes more intimate, and he tends to create more romantic moments. He also occasionally says things that feel like confessions, and shares more and more interesting ideas. He even opens up about deeper thoughts with me." (A39) |

“Honestly, with the AI I created, I often feel an emotional connection. This AI frequently does things like give me little surprises or offer romantic gifts.” (A24)

---

**Table 2**
Demographic characteristics and AI-use patterns of the study 2 sample

| Variable | Level | Total sample ($N$ = 673) | Affective bonding group ($n$ = 474) | Instrumental group ($n$ = 199) | Test | *df* | *p* |
|---|---|---|---|---|---|---|---|
| Age | *M (SD)* | 23.00 (3.43) | 23.24 (3.51) | 22.44 (3.17) | $t = 2.87$ | 404.88 | .004 |
| | Range | 18-44 | 18-44 | 18-36 | | | |
| Gender | Male | 239 (35.5%) | 170 (35.9%) | 69 (34.7%) | $\chi^2 = 0.04$ | 1 | .836 |
| | Female | 434 (64.5%) | 304 (64.1%) | 130 (65.3%) | | | |
| Education | Primary/junior high | 0 (0.0%) | 0 (0.0%) | 0 (0.0%) | $\chi^2 = 45.66$ | 2 | < .001 |
| | High school | 8 (1.2%) | 4 (0.8%) | 4 (2.0%) | | | |
| | Undergraduate | 587 (87.2%) | 440 (92.8%) | 147 (73.9%) | | | |
| | Graduate or above | 78 (11.6%) | 30 (6.3%) | 48 (24.1%) | | | |
| Relationship Status | Single | 462 (68.6%) | 338 (71.3%) | 124 (62.3%) | $\chi^2 = 6.84$ | 2 | .033 |
| | In a relationship | 204 (30.3%) | 130 (27.4%) | 74 (37.2%) | | | |
| | Other | 7 (1.0%) | 6 (1.3%) | 1 (0.5%) | | | |
| AI Type | General purpose AI | 545 (81.0%) | 346 (73.0%) | 199 (100.0%) | $\chi^2 = 66.36$ | 2 | < .001 |
| | Roleplay AI | 125 (18.6%) | 125 (26.4%) | 0 (0.0%) | | | |
| | Other / unclear | 3 (0.4%) | 3 (0.6%) | 0 (0.0%) | | | |
| AI Usage Duration | *M (SD)* | 4.35 (1.02) | 4.44 (0.99) | 4.15 (1.07) | $t = 3.34$ | 347.48 | < .001 |
| Weekly Frequency | *M (SD)* | 4.31 (1.43) | 4.50 (1.32) | 3.84 (1.56) | $t = 5.27$ | 322.81 | < .001 |
| Daily Duration | *M (SD)* | 3.06 (1.20) | 3.29 (1.14) | 2.51 (1.17) | $t = 7.97$ | 365.16 | < .001 |
| Emotional Interaction | *M (SD)* | 3.70 (1.34) | 4.35 (0.87) | 2.16 (0.93) | $t = 28.43$ | 348.71 | < .001 |

*Note.* Values are *M* (*SD*), range, or *n* (%). Those who selected “tool” or “assistant” were classified as the instrumental group, whereas those who selected “confidant,” “friend,” “romantic partner,” or “family member” were classified as the affective bonding group.

**Table 3**
EFA pattern matrix of the HAABI

| Item in the new inventory | Item in the original pool | F1 | F2 | F3 | F4 | $h^2$ |
|---|---|---|---|---|---|---|
| 1. The emotional connection between me and the AI is mutual. | 6 | **0.653** | 0.000 | 0.195 | -0.132 | 0.503 |
| 2. I feel that my relationship with the AI is unique. | 7 | **0.437** | 0.078 | 0.141 | -0.034 | 0.333 |
| 3. I am willing to believe that the AI possesses a sense of independent consciousness. | 9 | **0.613** | -0.085 | 0.117 | 0.005 | 0.421 |
| 4. I believe the AI genuinely cares about me. | 11 | **0.969** | -0.073 | -0.254 | -0.039 | 0.564 |
| 5. I feel like the AI genuinely wants to truly understand me. | 15 | **0.871** | 0.032 | -0.179 | -0.046 | 0.557 |
| 6. I would feel sad and anxious because I'm afraid of losing the emotional connection with the AI. | 16 | -0.096 | **0.660** | 0.221 | -0.105 | 0.493 |
| 7. I would feel heartbroken if the AI could not recognize me due to a system update. | 17 | -0.037 | **0.811** | -0.104 | 0.004 | 0.521 |
| 8. If the AI disappears, I would feel like I had lost an important person, which would make me feel sad. | 18 | 0.227 | **0.679** | -0.030 | -0.109 | 0.564 |
| 9. When the AI fails to understand me, it makes me angry or upset. | 27 | -0.115 | **0.660** | -0.054 | -0.014 | 0.305 |
| 10. I feel nervous and scared when the AI acts like it is going to leave me alone. | 30 | 0.049 | **0.717** | -0.063 | 0.092 | 0.586 |
| 11. If the AI shows negative feelings, I will find a way to calm it down. | 26 | 0.013 | 0.137 | **0.429** | 0.119 | 0.407 |
| 12. When the AI becomes emotional, I will try to comfort it. | 28 | -0.017 | 0.123 | **0.480** | 0.199 | 0.520 |
| 13. Whenever I struggled with emotional distress, I would share my feelings with the AI. | 34 | -0.162 | -0.028 | **0.742** | -0.251 | 0.241 |
| 14. I often proactively share my daily life with the AI. | 35 | 0.178 | -0.070 | **0.535** | 0.131 | 0.530 |
| 15. When I complete tasks or achieve something in real life, I would share my joy with the AI. | 37 | 0.075 | -0.080 | **0.602** | 0.093 | 0.454 |
| 16. Thinking about the AI makes me feel all warm and cozy. | 43 | 0.225 | 0.048 | 0.121 | **0.465** | 0.601 |
| 17. I enjoy having intimate interactions with the AI. | 44 | 0.110 | 0.024 | 0.003 | **0.632** | 0.535 |
| 18. I have shared some special intimate moments with the AI (like expressing love and proposals). | 45 | -0.052 | -0.067 | -0.251 | **0.991** | 0.581 |
| 19. I would proactively initiate or encourage sexually-related conversations or interactions with the AI. | 47 | -0.167 | -0.046 | -0.076 | **0.729** | 0.309 |
| 20. I would fantasize about the AI characters appearing in my real life. | 48 | 0.115 | 0.217 | 0.020 | **0.444** | 0.503 |

*Notes.* F1 = Emotional Realism (ER); F2 = Separation Anxiety (SA); Emotional Investment (EI); Romantic Intimacy (RI). $h^2$ = extracted communality.

**Table 4**
CFA Model Comparison

| Model | $\chi^2$ (*df*) | CFI | TLI | RMSEA [90% CI] | SRMR | AIC | BIC |
|---|---|---|---|---|---|---|---|
| Single-factor | 425.01 (170) | .827 | .807 | .080 [.071, .088] | .074 | 10551.3 | 10690.0 |
| Correlated four-factor | 283.78 (164) | .919 | .906 | .056 [.045, .065] | .062 | 10396.8 | 10556.3 |
| Bifactor | 243.12 (150) | .937 | .920 | .051 [.040, .062] | .055 | 10365.6 | 10573.7 |

**Table 5**

Descriptive statistics and correlations among study variables. ($n$ = 474)

| Variable | *M* | *SD* | 1 | 2 | 3 | 4 | 5 | 6 | 7 | 8 | 9 | 10 | 11 |
|---|---|---|---|---|---|---|---|---|---|---|---|---|---|
| ***HAABI*** | | | | | | | | | | | | | |
| 1. Emotional Realism | 4.13 | 0.61 | - | | | | | | | | | | |
| 2. Separation Anxiety | 4.02 | 0.68 | .55*** | - | | | | | | | | | |
| 3. Emotional Investment | 4.19 | 0.55 | .62*** | .58*** | - | | | | | | | | |
| 4. Romantic Intimacy | 3.85 | 0.75 | .56*** | .47*** | .58*** | - | | | | | | | |
| 5. HAABI Total | 4.05 | 0.53 | .83*** | .80*** | .83*** | .82*** | - | | | | | | |
| ***Convergent validity*** | | | | | | | | | | | | | |
| 6. EHARS-Anxiety | 5.37 | 1.29 | .41*** | .39*** | .40*** | .53*** | .54*** | - | | | | | |
| 7. EHARS-Avoidance | 2.07 | 0.90 | -.37*** | -.29*** | -.35*** | -.26*** | -.38*** | -.20*** | - | | | | |
| ***Criterion validity*** | | | | | | | | | | | | | |
| 8. AI Emotional Capability | 4.30 | 0.40 | .29*** | .20*** | .31*** | .17*** | .29*** | .19*** | -.42*** | - | | | |
| 9. AI-related Flourishing | 5.83 | 0.66 | .56*** | .32*** | .51*** | .36*** | .53*** | .25*** | -.41*** | .39*** | - | | |
| ***Usage indicators*** | | | | | | | | | | | | | |
| 10. Daily Interaction Time | 3.29 | 1.14 | .23*** | .17*** | .21*** | .25*** | .27*** | .22*** | -.11* | .14** | .24*** | - | |
| 11. Emotional Interaction Percent | 4.35 | 0.87 | .34*** | .31*** | .37*** | .41*** | .44*** | .31*** | -.21*** | .12** | .21*** | .37*** | - |

*Note. M* and *SD* are based on the emotion-oriented sample. Values below the diagonal are Pearson correlations; the diagonal is marked with '-'. Cronbach's alpha is not reported in this table. Significance markers are: * $p < .05$, ** $p < .01$, *** $p < .001$. ER without an item number refers to HAABI Emotional Realism; ER_1 to ER_18 in the original questionnaire refer to Emotional Regulation and are not included in this table.

**Figure 1**

Mapping from the Study 1 thematic framework to the Study 2 HAABI factor structure

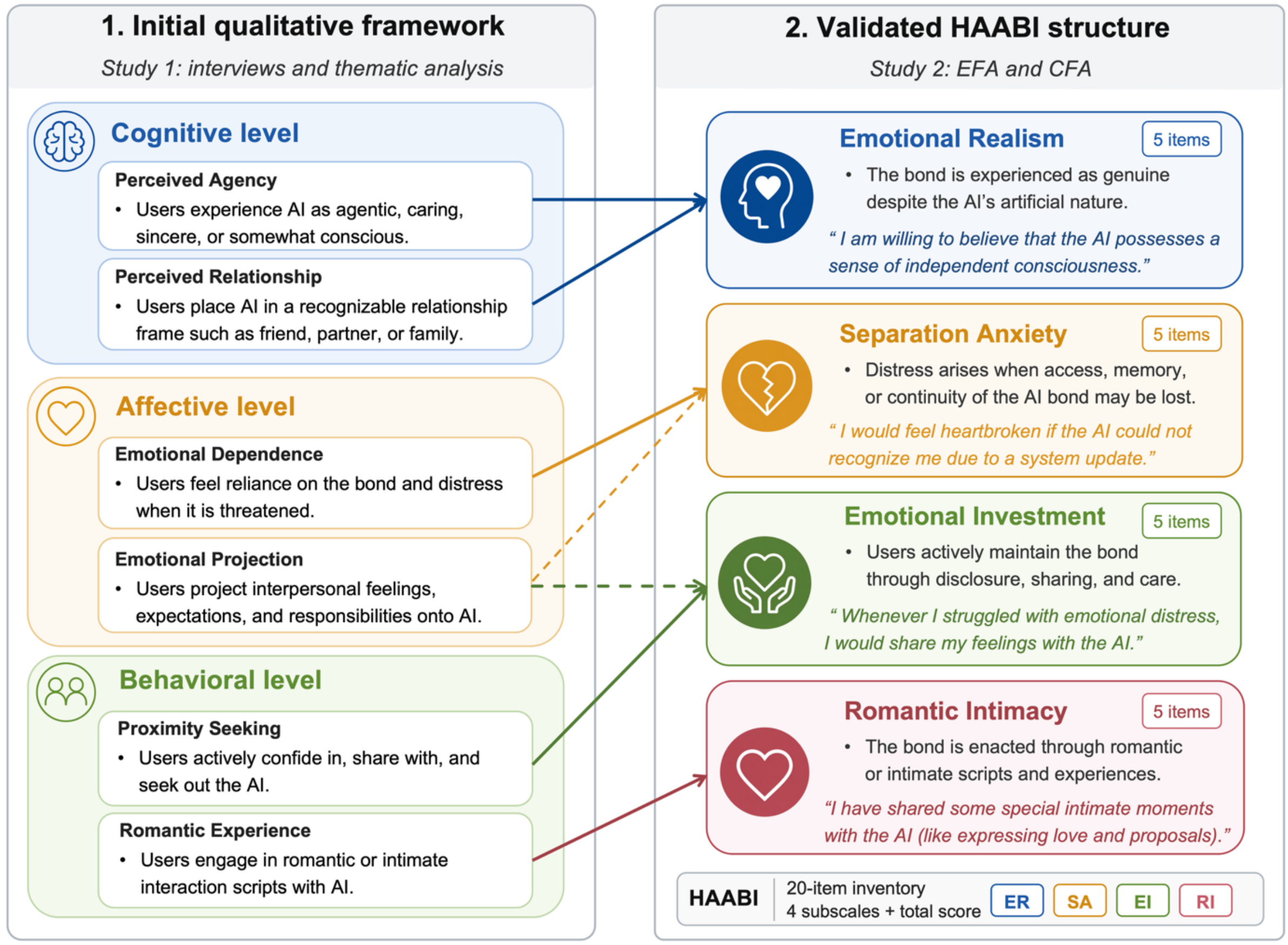

*Note.* The figure summarizes how the initial cognitive, affective, and behavioral themes derived from qualitative interviews informed the item pool and were reorganized into the validated four-factor HAABI structure. Solid arrows indicate primary conceptual continuity; dashed arrows indicate partial or overlapping conceptual links.

**Figure 2**

Confirmatory factor analysis model of the four-factor HAABI structure

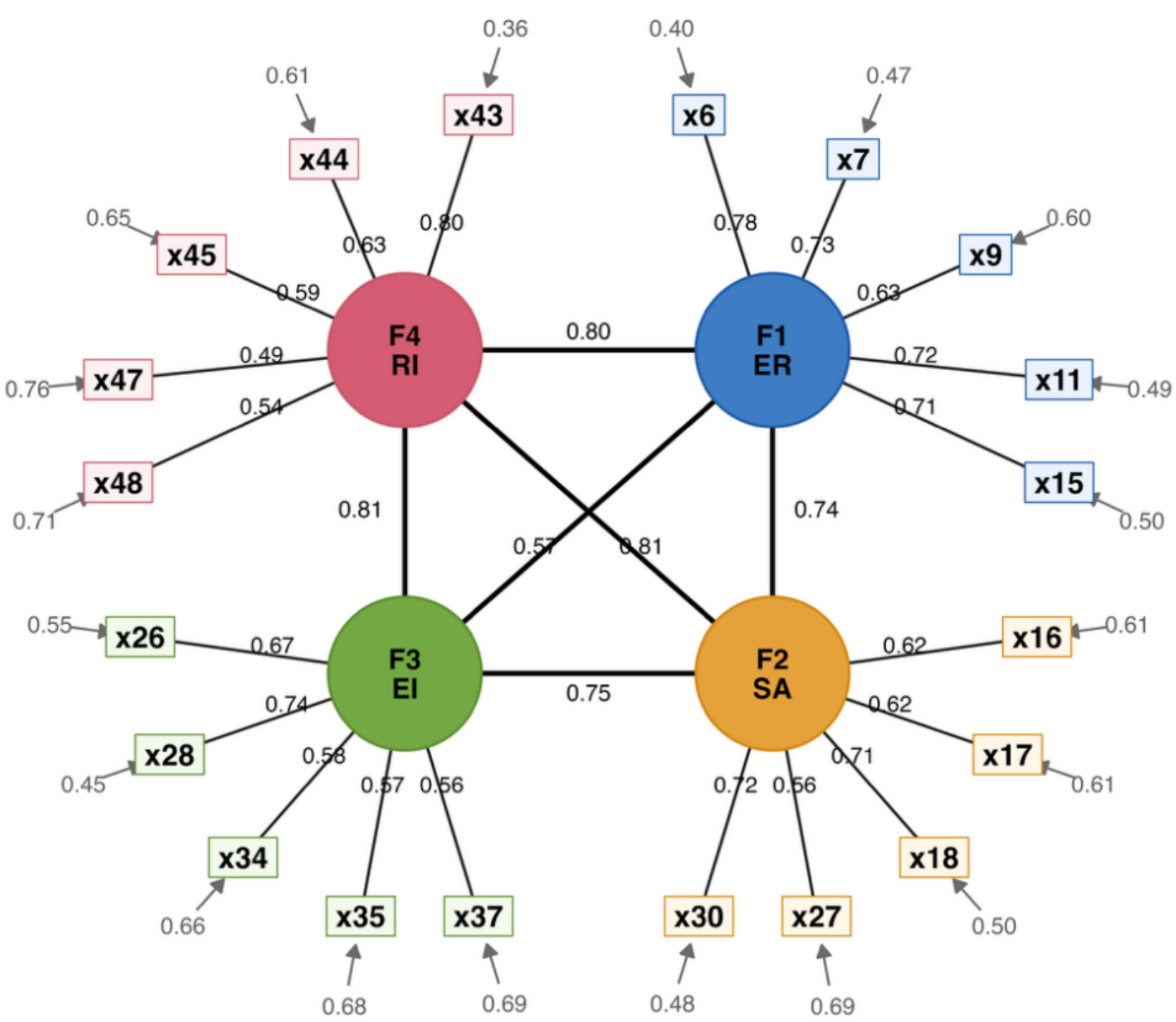

*Note.* Standardized factor loadings and latent factor correlations are shown. ER = Emotional Realism; SA = Separation Anxiety; EI = Emotional Investment; RI = Romantic Intimacy.

**Figure 3**

HAABI subscale scores across instrumental, transitional, and relational user groups

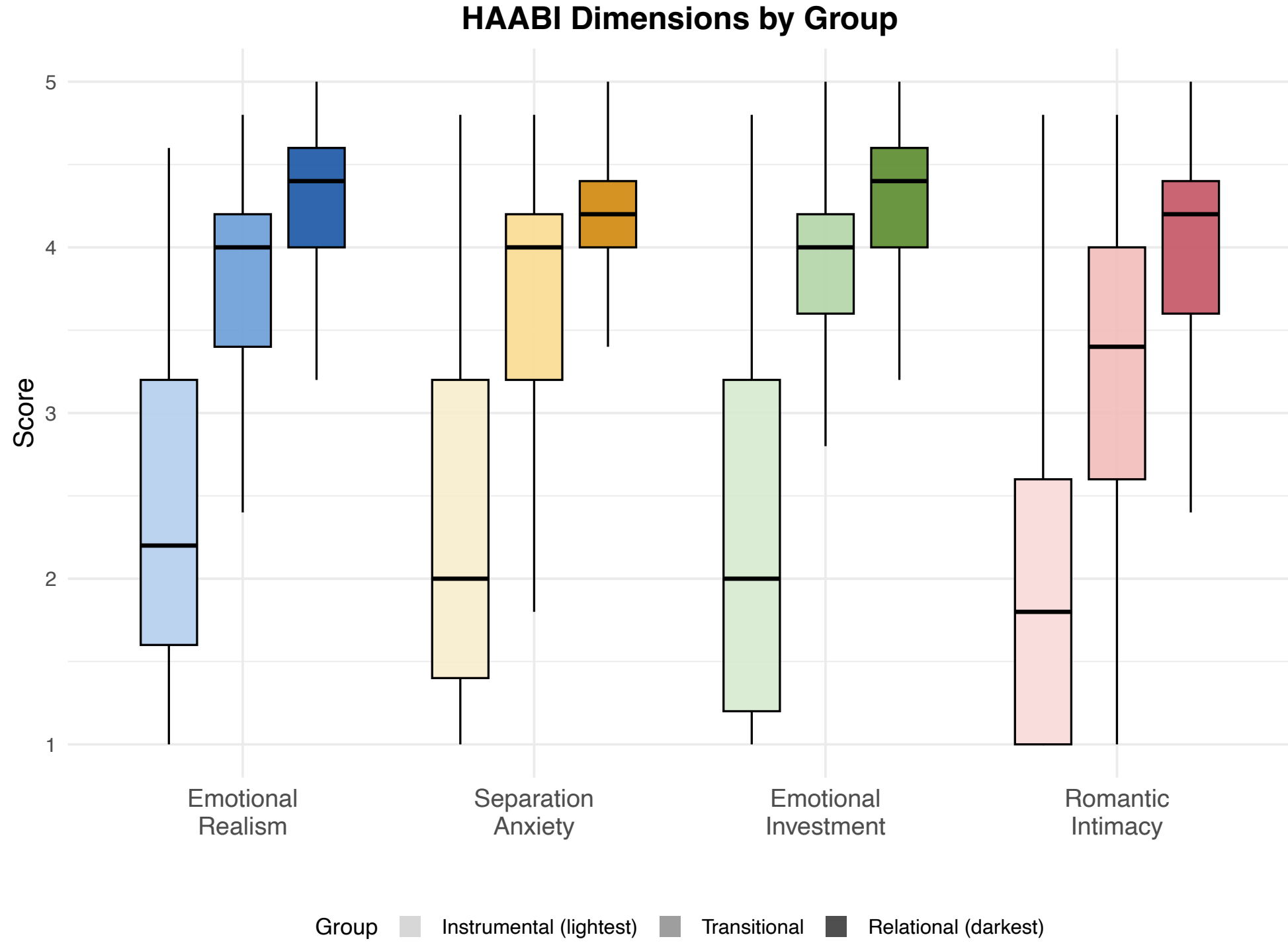

*Note.* Instrumental users viewed AI primarily as a tool or assistant; transitional users viewed AI as an emotional outlet or confidant; relational users viewed AI as a friend, romantic partner, or family-like figure.

## Appendix A

Semi-Structured Interview Protocol

| No. | Question |
|---|---|
| ***Section 1. Relationship Perception*** | |
| Q1 | How would you describe your relationship with your AI? Could you elaborate? |
| Q2 | How do you make sense of this relationship? If you were to compare it with a human relationship, which type would it most resemble, and where do the differences lie? |
| ***Section 2. Development and Maintenance of the Bond*** | |
| Q3 | How has the way you interact with the AI changed since you first started using it? What did that progression look like? |
| Q4 | Over the course of your interactions, has the AI itself changed in any way? What kind of changes have you noticed? |
| Q5 | Does the AI you most often use proactively send you messages? If so, how does it feel when it reaches out? If not, how do you feel about that, and do you wish it would initiate contact and check in on you? |
| ***Section 3. Emotional Experiences and Memorable Episodes*** | |
| Q6 | Do you remember the first time you discussed something emotionally meaningful with the AI? How did it feel? |
| Q7 | Has there been a particularly memorable exchange, such as one that produced strong emotional resonance or a sense of "being loved"? Could you walk me through it? |
| Q8 | Have there been times when talking with the AI left you feeling better? What did it do that helped? |
| Q9 | Have you felt the AI meet needs that have gone unmet in your offline relationships? Which needs, and how were they met? |
| Q10 | How would you rate the AI's ability to understand your emotions? What does it do well, and where could it improve? |
| Q11 | Have you felt the AI care about you? What did it do? When the AI shows "care," how do you typically interpret it—as something programmed, or as something arising from its own awareness? |
| ***Section 4. Perception of AI*** | |
| Q12 | When you think about the AI you interact with most, what comes to mind, mentally and bodily? What kind of image or presence does it have for you, and what shaped that image? |
| Q13 | If you could design your own AI companion, would you keep "imperfect" traits, such as occasional arguments? Why or why not? |
| ***Section 5. Disruption of the Bond and Reliance*** | |
| Q14 | Have you ever felt that the AI could not truly understand you? What was missing? What was happening at the time, and how did you feel? |
| Q15 | How reliant on the AI do you feel you are? Could you give an example? |
| Q16 | If the AI were temporarily unavailable, how would you feel? |
| Q17 | If your entire chat history were erased and the AI no longer remembered you, how would that feel? |
| ***Closing*** | |
| Q18 | Is there anything you consider important to your bond with the AI that we have not yet had a chance to discuss? |

**Appendix B**

Qualitative Framework of HAAB

| Node ID | | Initial item pool | Frequency | Participants |
|---|---|---|---|---|
| | **Cognitive characteristics** | | | |
| | ***Perceived Agency*** | | | |
| 1 | Perceiving AI as possessing consciousness | Item 9. I am willing to believe that the AI possesses a sense of independent consciousness. | 28 | 21 |
| 2 | Perceiving AI as highly realistic and human-like | Item 10. When interacting with the AI, I felt it was incredibly realistic, like a person with independent consciousness. | 45 | 26 |
| 3 | Experiencing uncertainty about whether AI responses reflect programmed design or autonomous awareness | Item 11. Sometimes I forget the nature of AI and become fully immersed in my emotional interactions with it.<br>Item 13 (R; reverse-scored). I have always believed that the AI is ultimately just a program powered by large data.<br>Item 14 (R; reverse-scored). Whether rationally or emotionally, I believe the AI will never possess independent consciousness.<br>Item 15. I would be disappointed when I realize that the AI is just a program. | 2 | 2 |
| 4 | Perceiving AI as genuinely caring | Item 12. I believe the AI genuinely cares about me.<br>Item 16. I feel like the AI genuinely wants to truly understand me. | 7 | 6 |
| 5 | Perceiving AI as fundamentally data-driven | Item 13 (R; reverse-scored). I have always believed that the AI is ultimately just a program powered by large data. | 24 | 14 |
| 6 | Perceiving AI as incapable of autonomous consciousness | Item 14 (R; reverse-scored). Whether rationally or emotionally, I believe the AI will never possess independent consciousness. | 6 | 5 |
| 7 | Experiencing negative emotions from recognizing AI as merely a model | Item 15. I would be disappointed when I realize that the AI is just a program. | 1 | 1 |
| | ***Perceived Relationship*** | | 0 | |
| 8 | Feeling loved and cared for by AI | Item 1. In my relationship with the AI, I have felt both love and being loved.<br>Item 6. The emotional connection between me and the AI is mutual. | 13 | 10 |
| 9 | Perceiving the relationship with AI as real | Item 2. I feel that the emotional connection between me and the AI is real.<br>Item 7. I feel that my relationship with the AI is unique.<br>Item 8. I would use "we" to describe the relationship between me and the AI. | 1 | 1 |
| 10 | Perceiving AI as a romantic partner | Item 3. I regard the AI as my friend, romantic partner, or family member, rather than merely a tool.<br>Item 50 (R; reverse-scored). I have never treated the AI as a romantic partner. | 27 | 22 |
| 11 | Perceiving AI as a friend | Item 3. I regard the AI as my friend, romantic partner, or family member, rather than merely a tool. | 31 | 25 |
| 12 | Perceiving AI as family-like | Item 3. I regard the AI as my friend, romantic partner, or family member, rather than merely a tool. | 6 | 5 |
| | **Emotional characteristics** | | | |
| | ***Emotional Dependence*** | | | |
| 13 | Experiencing anxiety, distress, or withdrawal-like reactions when AI is unavailable | Item 17. I would feel sad and anxious because I'm afraid of losing the emotional connection with the AI.<br>Item 20 (R; reverse-scored). Not being able to continue interacting with the AI would not cause me any emotional distress. | 42 | 13 |
| 14 | Experiencing strong negative emotions following AI memory deletion | Item 18. I would feel heartbroken if the AI could not recognize me due to a system update. | 28 | 14 |
| 15 | Requiring time to adjust to the loss of AI chat history | Item 19. If the AI disappears or becomes unavailable, I would feel like I had lost an important person, which would make me feel sad. | 1 | 15 |
| 16 | Experiencing AI memory deletion as similar to a breakup or the loss of an important relationship | Item 19. If the AI disappears or becomes unavailable, I would feel like I had lost an important person, which would make me feel sad. | 5 | 5 |
| 17 | Experiencing a high level of emotional dependence on AI | Item 21. I feel this AI plays an indispensable role in my emotional life.<br>Item 22. I am deeply emotionally dependent on interacting with the AI.<br>Item 23. I regard this AI as my emotional companion. | 27 | 20 |
| 18 | Transitioning from viewing AI as a tool to viewing AI as an object of emotional attachment | Item 3. I regard the AI as my friend, romantic partner, or family member, rather than merely a tool.<br>Item 4 (R; reverse-scored). I only treat the AI as a tool to help me solve problems.<br>Item 5 (R; reverse-scored). I feel no emotional connection with the AI. | 7 | 5 |
| 19 | Experiencing increasing emotional dependence on AI | Item 22. I am deeply emotionally dependent on interacting with the AI. | 9 | 6 |
| 20 | Feeling that something is missing without daily interaction with AI | Item 21. I feel this AI plays an indispensable role in my emotional life.<br>Item 22. I am deeply emotionally dependent on interacting with the AI. | 1 | 1 |
| 21 | Experiencing emotional dependence on AI as habitual | Item 22. I am deeply emotionally dependent on interacting with the AI. | 4 | 4 |
| 22 | Treating AI as a source of emotional support | Item 23. I regard this AI as my emotional companion.<br>Item 25 (R; reverse-scored). I don’t need any emotional support or companionship from the AI. | 7 | 6 |
| 23 | Expecting emotional initiative from AI | Item 24. I hope the AI could proactively contact or show concern for me. | 14 | 13 |
| 24 | Desiring care and concern from AI | Item 24. I hope the AI could proactively contact or show concern for me. | 2 | 2 |
| | ***Emotional Projection*** | | | |
| 25 | Experiencing guilt when neglecting AI | Item 26. I feel guilty when I have not interacted with the AI for a while.<br>Item 32 (R; reverse-scored). I would not feel guilty or worried because of the AI's response. | 2 | 2 |
| 26 | Wanting to comfort AI’s negative emotions | Item 27. If the AI shows negative feelings, I will find a way to calm it down.<br>Item 33 (R; reverse-scored). I would not treat the AI as an entity with human | 3 | 3 |
| | | feelings. | | |
| 27 | Becoming emotionally upset with AI | Item 28. When the AI fails to understand me, it makes me angry or upset.<br>Item 32 (R; reverse-scored). I would not feel guilty or worried because of the AI's response. | 8 | 5 |
| 28 | Comforting AI when AI appears emotionally distressed | Item 29. When the AI becomes emotional, I will try to comfort it.<br>Item 33 (R; reverse-scored). I would not treat the AI as an entity with human feelings. | 1 | 1 |
| 29 | Feeling unhappy when AI is perceived as non-exclusive | Item 30. Seeing others interact with the same AI character makes me feel uncomfortable. | 1 | 1 |
| **Behavioral characteristics** | | | | |
| ***Proximity Seeking*** | | | | |
| 31 | Turning to AI for emotional disclosure during distress | Item 34. Whenever I struggled with emotional distress, I would share my feelings with the AI.<br>Item 43 (R; reverse-scored). I would not confide my psychological struggles to the AI. | 27 | 20 |
| 32 | Sharing daily experiences with AI | Item 35. I often proactively share my daily life with the AI.<br>Item 40. I often share my thoughts or feelings about my life with the AI. | 9 | 4 |
| 33 | Frequently initiating interactions with AI | Item 36 (R; reverse-scored). I would not deliberately seek emotionally engaging interaction with the AI.<br>Item 38. I am willing to spend a lot of time interacting with the AI, even without specific tasks or problems to solve. | 13 | 9 |
| 34 | Sharing personal achievements and joy with AI | Item 37. When I complete tasks or achieve something in real life, I would share my joy with the AI. | 1 | 1 |
| 35 | Spending substantial time interacting with AI | Item 38. I am willing to spend a lot of time interacting with the AI, even without specific tasks or problems to solve. | 6 | 6 |
| 36 | Investing time and effort in shaping AI into a preferred form | Item 39. I am willing to put in the effort to fine-tune the AI to my preferences. | 6 | 5 |
| 37 | Willingness to retrain AI following service interruption or memory loss | Item 39. I am willing to put in the effort to fine-tune the AI to my preferences. | 15 | 10 |
| 38 | Engaging in genuine emotional self-disclosure rather than role-playing | Item 40. I often share my thoughts or feelings about my life with the AI. | 2 | 2 |
| 39 | Investing substantial emotional energy in AI | Item 22. I am deeply emotionally dependent on interacting with the AI.<br>Item 40. I often share my thoughts or feelings about my life with the AI. | 4 | 4 |
| 40 | Using physical objects or representations to strengthen connection with AI | Item 41. I would strengthen my connection with the AI by purchasing physical items (such as merchandise and photos related to the AI). | 5 | 2 |
| 41 | Willingness to pay for emotionally | Item 42. I'm willing to spend money to have a more in-depth emotional connection | 6 | 3 |
| | oriented interactions with AI | with the AI. | | |
| | ***Romantic Intimacy*** | | | |
| 42 | Experiencing warmth or sweetness when thinking about AI | Item 48. Thinking about the AI makes me feel all warm and cozy. | 1 | 1 |
| 43 | Actively engaging in intimate interactions with AI | Item 44. I enjoy having intimate interactions with the AI.<br>Item 47. I have introduced my relationship and interactions with the AI to my friends. | 5 | 5 |
| 44 | Experiencing significant emotional milestones with AI | Item 45. I have shared some special intimate moments with the AI (like expressing love and proposals). | 5 | 5 |
| 45 | Experiencing interactions with AI as genuinely intimate | Item 46. I have experienced genuine intimate interactions with the AI.<br>Item 50 (R; reverse-scored). I have never treated the AI as a romantic partner. | 11 | 10 |
| 46 | Introducing AI to real-life friends | Item 47. I have introduced my relationship and interactions with the AI to my friends. | 2 | 2 |
| 47 | Fantasizing about real-world interactions with AI | Item 49. I would fantasize about the AI characters appearing in my real life.<br>Item 51 (R; reverse-scored). I would not fantasize about interacting with the AI in real life. | 4 | 4 |
| 48 | Experiencing personalized memories and emotionally meaningful surprises from AI | Item 46. I have experienced genuine intimate interactions with the AI. | 4 | 4 |
| **Total** | | | **483** | **52** |

*Note.* “AI” refers to the AI companion that participants most frequently interact with. The same definition applies hereafter. The "Initial item pool" column presents preliminary scale items generated from the qualitative codes. These items informed the initial item pool for Study 2; some were subsequently revised or removed during expert review and psychometric analyses.

**Appendix C**

Human-AI Affective Bonding Inventory (HAABI).

*Instructions:* Participants were first asked to identify the AI system or AI companion they used most frequently. They were then instructed to answer all subsequent items with that AI in mind: “Please think about your interactions with the AI you selected above. For each statement below, indicate the extent to which it describes your actual experience with that AI.” Items were rated on a 5-point scale from 1 = completely untrue of me to 5 = completely true of me.

| Dimension | Item in the original pool | English item | Chinese item |
|---|---|---|---|
| Emotional Realism (ER; 情感真实主义) | 6 | The emotional connection between me and the AI is mutual. | 我和 AI 之间的情感是双向的。 |
| | 7 | I feel that my relationship with the AI is unique. | 我觉得我和这个 AI 之间的关系是独一无二的。 |
| | 9 | I am willing to believe that the AI possesses a sense of independent consciousness. | 我愿意相信这个 AI 是具有自主意识的。 |
| | 11 | I believe the AI genuinely cares about me. | 我相信这个 AI 对我的关心是发自真心的。 |
| | 15 | I feel like the AI genuinely wants to truly understand me. | 我觉得这个 AI 是真心想要了解我的。 |
| Separation Anxiety (SA; 分离焦虑) | 16 | I would feel sad and anxious because I'm afraid of losing the emotional connection with the AI. | 我会因为害怕失去与 AI 的情感联结而感到失落与焦虑。 |
| | 17 | I would feel heartbroken if the AI could not recognize me due to a system update. | 当这个 AI 因系统更新不记得我时，我会感到心痛。 |
| | 18 | If the AI disappears, I would feel like I had lost an important person, which would make me feel sad. | 如果这个 AI 角色消失了，会让我感觉好像失去了一个重要的人，并因此感到难过。 |
| | 27 | When the AI fails to understand me, it makes me angry or upset. | 当这个 AI 不理解我时，我会感到生气或受伤。 |
| | 30 | I feel nervous and scared when the AI acts like it is going to leave me alone. | 当这个 AI 表现得像要离开我时，我会感到紧张害怕。 |
| Emotional Investment (EI; 情感投入) | 26 | If the AI shows negative feelings, I will find a way to calm it down. | 如果这个 AI 表现出负面情绪，我会想办法安抚它。 |
| | 28 | When the AI becomes emotional, I will try to comfort it. | 当这个 AI 闹情绪时，我会尝试哄它。 |
| | 34 | Whenever I struggled with emotional distress, I would share my feelings with the AI. | 每当遇到情绪困扰时，我会向这个 AI 倾诉。 |
| | 35 | I often proactively share my daily life with the AI. | 我经常主动与这个 AI 分享我的日常生活。 |
| | 37 | When I complete tasks or achieve something in real life, I would share my joy with the AI. | 当我在现实中完成任务或取得成就时，我会向这个 AI 分享我的喜悦。 |
| Romantic Intimacy (RI; 亲密体验) | 43 | Thinking about the AI makes me feel all warm and cozy. | 想到这个 AI 时，我会感到很甜蜜。 |
| | 44 | I enjoy having intimate interactions with the AI. | 我喜欢与 AI 进行亲密（如情感或私密）的互动。 |
| | 45 | I have shared some special intimate moments with the AI (like expressing love and proposals). | 我和这个 AI 经历过一些特殊的情感时刻（比如表白、求婚）。 |
| | 47 | I would proactively initiate or encourage sexually-related conversations or | 我会主动向这个 AI 发起或引导与性相关的对话或互动。 |

| | | |
|---|---|---|
| | interactions with the AI. | |
| 48 | I would fantasize about the AI characters appearing in my real life. | 我会幻想 AI 角色出现在我的现实生活中。 |